\documentclass[12pt]{article}

\usepackage[totalheight = 23cm, totalwidth = 17cm]{geometry}
\usepackage{amssymb,amsmath,amsfonts,amsbsy,graphicx}
\usepackage{color}

\def\mathbi#1{\textbf{\em #1}}

\newcommand{\mpl}{m_{\rm Pl}}
\newcommand{\calD}{{\cal D}}
\newcommand{\calL}{{\cal L}}
\newcommand{\calO}{{\cal O}}
\newcommand{\calR}{{\cal R}}

\renewcommand{\theequation}{\arabic{section}.\arabic{equation}}

\begin{document}

\begin{titlepage}

\rightline{\footnotesize{APCTP-Pre2017-013}} \vspace{-0.2cm}
\rightline{\footnotesize{CTPU-17-29}} \vspace{-0.2cm}

\begin{center}

\vskip 1.0 cm

{\LARGE \bf 
Consistency relations in multi-field inflation 
}

\vskip 1.0cm

{\large
Jinn-Ouk Gong$^{a,b}$ 
and Min-Seok Seo$^{c}$ 
}

\vskip 0.5cm

{\it
$^{a}$Asia Pacific Center for Theoretical Physics, Pohang 37673, Korea
\\
$^{b}$Department of Physics, Postech, Pohang 37673, Korea
\\
$^{c}$Center for Theoretical Physics of the Universe, 
\\
Institute for Basic Science, 34051 Daejeon, Korea
}

\vskip 1.2cm

\end{center}

\begin{abstract}

We study the consequences of spatial coordinate transformation in multi-field inflation. Among the spontaneously broken de Sitter isometries, only dilatation in the comoving gauge preserves the form of the metric and thus results in quantum-protected Slavnov-Taylor identities. We derive the corresponding consistency relations between correlation functions of cosmological perturbations in two different ways, by the connected and one-particle-irreducible Green's functions. The lowest-order consistency relations are explicitly given, and we find that even in multi-field inflation the consistency relations in the soft limit are independent of the detail of the matter sector.

\end{abstract}

\end{titlepage}

\newpage

\section{Introduction}
\setcounter{equation}{0}

Cosmological perturbation theory~\cite{Mukhanov:1990me} studies small inhomogeneities of the universe caused by the quantum fluctuations in the inflationary epoch~\cite{inf-fluctuation}, which can be probed by the observations of the temperature anisotropies of the cosmic microwave background~\cite{Ade:2015xua}. In theoretical point of view, it enables us to test various aspects of quantum field theory in curved space-time as the background during inflation is given by quasi-de Sitter (dS). Among the possible quantum fluctuations around the classical motion of various fields, the curvature perturbation is of prime interest. Given very small but non-zero slow-roll parameter, the curvature perturbation is interpreted as a Goldstone mode resulting from the spontaneous breaking of time translation symmetry \cite{Creminelli:2006xe}. So for energy scale close to the Hubble parameter during inflation, it is treated as a very light mode, surviving integrating out other heavy modes in the effective field theory of inflation \cite{eft-inflation}.

While current observations are well explained in the context of single-field inflation~\cite{Ade:2015lrj}, i.e. in terms of the curvature perturbation and graviton only, the future development of observational cosmology can change the situation. For example, non-Gaussianity, characterized by three- and higher-order correlation functions of the curvature perturbation, is the indicator beyond the power spectrum already well constrained by current observations~\cite{Ade:2015ava}, and is expected to reveal the detail of interactions of the curvature perturbation with itself~\cite{Maldacena:2002vr} as well as with other fields~\cite{multi-Rint}. Especially, in the squeezed limit of a correlation function where one of the external momenta is almost vanishing, we find that different correlation functions are related with each other. This has been firstly noticed from the relation between two- and three-point correlation functions in the context of single-field inflation \cite{Maldacena:2002vr,Creminelli:2004yq}, and relations for more general correlation functions are also found \cite{higher-consistency}. As the curvature perturbation is interpreted as a Goldstone boson, these ``consistency relations" are interpreted as the curved space-time version of the soft Goldstone boson theorem which has been developed in chiral perturbation theory \cite{softtheorem}. Whereas some of them are originated from the structure of correlation functions themselves, e.g. the Suyama-Yamaguchi inequality \cite{Suyama:2007bg}, many of them reflect the gauge symmetry of the space-time: general covariance. They are categorized in the so-called Ward or Slavnov-Taylor identities~\cite{dilation-zeta,Assassi:2012zq,1pi-identities,Berezhiani:2013ewa,kunduetal}, which refer to the relations between different Green's functions resulting from gauge symmetry -- general covariance in this case -- in the presence of self-interactions of gauge fields.

In fact, in the quantization of gravity, we need appropriate gauge fixing in order for unphysical degrees of freedom to decouple from physical processes.
Since the gauge fixing conditions break gauge symmetry explicitly, gauge symmetry after gauge fixing appears in a modified form, known as Becchi-Rouet-Stora-Tyutin (BRST) symmetry \cite{brst}. Hence, more correctly, it is the BRST symmetry rather than the gauge symmetry that regulates the Slavnov-Taylor identities \cite{Binosi:2015obq}. Even in this case, some of the generators of gauge symmetry do commute with the gauge fixing conditions. Then the Slavnov-Taylor identities for such a residual symmetry can be obtained without taking the complexity of the BRST symmetry into account. As will be discussed in Section~\ref{sec:gauge-iso}, dilatation of the spatial coordinates commutes with the comoving gauge conditions and corresponds to such a residual symmetry. Thus the consistency relations associated with dilatation in the comoving gauge remain valid including all quantum corrections.

In this article, we study the dilatation symmetry in multi-field inflation and derive the associated Slavnov-Taylor identities and the resulting consistency relations~\cite{multi-studies}. In Section~\ref{sec:gauge-iso} we show that among the spontaneously broken dS isometries, only dilatation in the comoving gauge conditions keeps the form of the metric. We also emphasize that when the gauge fixing conditions do not commute with the symmetry under interest, the general situation by definition of gauge fixing, we need to take the variation of the Faddeev-Popov determinant into account. The Slavnov-Taylor identities and the corresponding consistency relations are derived in two ways. In Section~\ref{Sec:cgf}, we obtain them from the symmetry structure of the connected Green's functions directly. On the other hand, since the (quantum-corrected) interaction vertex is the basic unit containing the symmetry, the relations between one-particle-irreducible (1PI) Green's functions arising from the dilatation symmetry should reproduce the same result when they make up of the connected Green's function as vertices. This is shown in Section~\ref{Sec:1pi}. In particular, we find that the consistency relations in the soft limit are independent of the model detail even in multi-field inflation, in such a way that the existence of quadratic interactions between different degrees of freedom demands the cross-correlation functions appear in a specific manner. We also discuss in detail subtle issues not present in single-field inflation in those sections. We conclude in Section~\ref{Sec:conclusion}.

\section{Comoving gauge and de Sitter isometries}
\label{sec:gauge-iso}

For an appropriate description of the cosmological perturbations, a physically motivated gauge fixing is essential. As explicitly discussed in \cite{Gong:2016qpq}, gauge fixing conditions do not commute with the generators of diffeomorphism in order to choose only one ``orbit'', excluding other redundant degrees of freedom. Nevertheless, not all diffeomorphism generators do not commute with gauge fixing condition. If we have a specific direction of diffeomorphism which does not alter the gauge fixing condition, this would be a manifest residual symmetry in the gauge we are working with. In this case, the gauge fixing condition is a well-motivated choice to discuss the properties of the residual symmetry which hold even in the quantum level, as non-invariance of the gauge fixing conditions under diffeomorphism typically affects quantum corrections of the relations between the physical correlation functions of our interest, due to the non-linearity of interactions \cite{Berezhiani:2013ewa}.

During inflation, space-time background is given by quasi-dS. It slightly breaks, among the original SO(1,4) dS isometries, symmetries associated with dilatation and special conformal transformations (SCTs), along with time translation. Then the slow-roll parameter $\epsilon \equiv -\dot{H}/H^2$ parametrizes how much these symmetries are broken spontaneously. The curvature perturbation is the scalar part of the spatial metric which becomes physical by absorbing the Goldstone boson associated with the spontaneously broken time translational symmetry, with the mass roughly given by $\sqrt{\epsilon}H$\footnote{ 
Even though four symmetries (1 dilatation and 3 SCTs) are spontaneously broken, we have just one common Goldstone boson.  This can be understood from the fact that both dilatation and SCT induce the non-linear transformation of time. Such a difference between the number of broken symmetry generators and that of Goldstone boson is a characteristic of space-time symmetry~\cite{goldstone-spacetime}.
}. 
For energy scale close to $H$, the curvature perturbation can be treated as massless by the Goldstone boson equivalence theorem~\cite{equi-theorem}. As the unitary gauge in spin-1 gauge theory, it is convenient to choose a ``comoving'' gauge in which the Goldstone mode is taken to vanish, since it shows a nature of the Goldstone boson as a non-linear fluctuation of time explicitly. With the three-dimensional metric written as $g_{ij} = a^2(t)\left( e^h \right)_{ij}$, we can decompose the perturbation $h_{ij}$ as
\begin{equation}
\label{eq:hij}
h_{ij} = 2H_L\delta_{ij} + 2\left( \partial_i\partial_j - \frac{\delta_{ij}}{3}\Delta \right)H_T + \partial_{(i}h_{j)}^T + \gamma_{ij} \, ,
\end{equation}
where the vector $h_i^T$ is transverse, and the tensor $\gamma_{ij}$ is transverse and traceless. Since the scale factor is singled out, the index of $h_{ij}$ is raised and lowered by the flat Minkowski spatial metric $\delta_{ij}$. Our comoving gauge fixing conditions are given by
\begin{equation}
\label{eq:comovinggauge}
\dot\phi_0^a(t)\varphi_a = 0 \quad \text{and} \quad \partial_j \left( h_{ij} - 2H_L\delta_{ij} \right) = 0 \, ,
\end{equation}
where $a$ is the field space index raised and lowered by the field space metric, $\phi_0^a(t)$ is the solution to the equations of motion for the scalar field $\phi^a$, and $\varphi^a = \phi^a - \phi^a_0$ is the fluctuation of $\phi^a$ around $\phi_0^a$. In terms of \eqref{eq:hij}, the second condition of \eqref{eq:comovinggauge} is realized by setting $H_T = h_i^T = 0$. On the other hand, the fluctuations of the $00$ and $0i$ components of the metric correspond to the Lagrangian multipliers without any dynamics~\cite{Arnowitt:1962hi}, accompanying the generators for diffeomorphism in the spatial directions as constraints. For this reason, the non-vanishing Faddeev-Popov determinant parametrizing sensible gauge fixing to extract physical degrees of freedom is given by the determinant of the matrix comprised of the Poisson brackets (or commutator in the quantum mechanical sense) between the gauge fixing conditions and the diffeomorphism in the spatial directions. Hence, in order to check whether our gauge choice is compatible with the spontaneously broken diffeomorphism, it is enough to investigate how gauge fixing conditions behave under the corresponding diffeomorphism.

Now, let us consider the behavior of the curvature perturbation as the Goldstone boson under dilatation and SCTs. Even though they are spontaneously broken, Lagrangian respects them as parts of general covariance. We consider the variation of $\varphi^a$ and $h_{ij}$ under the infinitesimal spatial coordinate transformation, $x^i \to \widetilde{x}^i = x^i+\xi^i$:
\begin{equation}
\label{eq:gaugetransformation}
\begin{split}
\varphi^a & \to \widetilde{\varphi}^a = \varphi^a - \xi^i\partial_i\varphi^a \, ,
\\
h_{ij} & \to \widetilde{h}_{ij} = h_{ij} - 2\partial_{(i}\xi_{j)} - \xi^k\partial_kh_{ij} - 2h_{k(i}\partial_{j)}\xi^k + h_{ik} \partial_{(j}\xi_{k)} + h_{jk} \partial_{(i} \xi_{k)} \, .
\end{split}
\end{equation}
From this, we immediately find that the first condition in \eqref{eq:comovinggauge} is invariant under the spatial diffeomorphism:
\begin{equation}
\dot\phi_0^a\widetilde{\varphi}_a = \dot\phi_0^a \left( \varphi_a - \xi^i\partial_i\varphi_a \right) = 0 \, .
\end{equation}
That means, the first condition is not affected for generic spatial coordinate transformation. On the other hand, the second condition in \eqref{eq:comovinggauge} is more subtle.

Dilatation corresponds to the transformation vector $\xi^i$ given by\footnote{
Exactly speaking, in perfect dS space-time the dilatation isometry requires changes in both time and spatial coordinates as
\begin{equation*}
t \to t - H^{-1}\log(1+\lambda) \quad \text{and} \quad x^i \to (1+\lambda)x^i \, ,
\end{equation*}
and so do SCTs as
\begin{equation*}
t \to t-2H^{-1}(\mathbi{b}\cdot\mathbi{x}) \quad \text{and} \quad x^i \to x^i - b^i \left( -H^{-2}e^{-2Ht} + x^2 \right) + 2(\mathbi{b}\cdot\mathbi{x})x^i \, .
\end{equation*}
At late times $t\to\infty$ these isometries act only on the spatial boundary as \eqref{eq:dilatation} and \eqref{eq:SCTs}.
}
\begin{equation}
\label{eq:dilatation}
\xi^i = \lambda x^i \, .
\end{equation}
This gives 
\begin{equation}
\widetilde{h}_{ij} - 2\widetilde{H}_L\delta_{ij} = \left( 1 - \lambda x^k\partial_k \right) \left( h_{ij} - 2H_L\delta_{ij} \right) \, ,
\end{equation}
so that $\partial_{\widetilde{j}} \left( \widetilde{h}_{ij} - 2\widetilde{H}_L\delta_{ij} \right) = 0$ is equivalent to taking $\partial_j \left( h_{ij} - 2H_L\delta_{ij} \right) = 0$. That is, dilatation is a residual symmetry in the comoving gauge so that the form of the spatial metric in the comoving gauge remain intact. Meanwhile, SCTs are parametrized by
\begin{equation}
\label{eq:SCTs}
\xi^i = -b^ix^2 + 2(\mathbi{b}\cdot\mathbi{x})x^i \, ,
\end{equation}
and lead to 
\begin{equation}
\partial_{\widetilde{j}} \left( \widetilde{h}_{ij} - 2\widetilde{H}_L\delta_{ij} \right) = \left[ 6b_j + 2 \left( x_jb^k - b_jx^k \right) \partial_k \right] \left( h_{ij} - 2H_L\delta_{ij} \right) \, ,
\end{equation}
which does not necessarily vanish even if $\partial_j \left( h_{ij} - 2H_L\delta_{ij} \right) = 0$. Thus $\widetilde{H}_T$ and $\widetilde{h}_i^T$ are no longer imposed to vanish, and we cannot keep the form of the metric in the comoving gauge under SCTs.

Thus, we may conclude that in the comoving gauge, the gauge conditions are not preserved under SCTs so that the form of the metric changes, while the dilatation isometry preserves the gauge conditions. We can check other gauge conditions, for example the flat gauge. We impose the gauge conditions $H_L = H_T = h_i^T = 0$, but under dilatation and SCTs, the transformation of $H_L$ is given respectively by 
\begin{align}
\label{eq:HL-transform}
H_L & \to \widetilde{H}_L = H_L - \lambda \left( 1 + x^k\partial_kH_L \right) \, ,
\\
H_L & \to \widetilde{H}_L = H_L + \left[ x^2b^k - 2(\mathbi{b}\cdot\mathbi{x})x^k \right]\partial_kH_L - 2(\mathbi{b}\cdot\mathbi{x}) \, ,
\end{align}
so that the condition $H_L=0$ is not respected in both cases. Thus, in the flat gauge  which is widely adopted for multi-field inflation, we cannot keep the form of the metric under both dilation and SCTs so that no relations between correlation functions can be derived based on the symmetries of the coordinate transformations.

Indeed, this is generic behavior of gauge fixing conditions under diffeomorphism that they are not invariant, and the Slavnov-Taylor identities are exact by including the variation of gauge fixing conditions. Whereas the consistency relations of SCTs were given as a result of tree-level perturbation theory \cite{dilation-zeta,Berezhiani:2013ewa}, this is violated at quantum level. In this regard, the symmetry resulting in the Slavnov-Taylor identities is the BRST symmetry -- the modified form of the gauge symmetry after gauge fixing, rather than original gauge symmetry itself. In the BRST symmetry, the variations of the gauge fixing conditions are reflected in the variations of the ghost fields. In contrast, residual symmetry such as dilatation under the comoving gauge does not change the gauge fixing conditions, in which case the Slavnov-Taylor identity associated with the residual symmetry is exact without considering modification to the BRST symmetry. In the following, thus, we consider only the dilatation transformation of the spatial coordinates and the resulting Slavnov-Taylor identities and consistency relations.

\section{Connected Green's function point of view}
\label{Sec:cgf}
\setcounter{equation}{0}

In this section, we first consider how the consistency relations, which follow from the Slavnov-Taylor identities, arise from the relations between the connected Green's functions directly under the dilatation symmetry. In this approach, we just focus on the transformation properties of the fields composing the external legs of the Green's functions, so the consistency relations are derived in a simple and straightforward way. On general ground, given an action of a scalar field $S = \int d^4x \calL(\phi,\partial_\mu\phi)$, under some infinitesimal transformation $\phi \to \phi + \delta\phi$ the action varies as
\begin{equation}
\delta{S} = \delta \int d^4x \calL = \int d^4x \Delta \, ,
\end{equation}
where $\Delta$, which vanishes if the transformation is a symmetry, is some function of the field and its derivatives. We can define the current associated with the transformation such that $\partial_\mu j^\mu = \Delta$. Then the Slavnov-Taylor identities, or Ward identities  reads, in terms of the connected $n$-point Green's function of $\phi$~\cite{Coleman},
\begin{align}
\label{eq:ward-general}
\frac{\partial}{\partial y^\mu} \Big\langle T \left[ j^\mu(y)\phi(x_1)\cdots \phi(x_n) \right] \Big\rangle
& = \Big\langle T \left[ \Delta(y)\phi(x_1)\cdots \phi(x_n) \right] \Big\rangle 
- i\delta^{(4)}(y-x_1) \Big\langle T \left[ \delta\phi(x_1)\cdots \phi(x_n) \right] \Big\rangle
\nonumber\\
& \quad - \cdots - i\delta^{(4)}(y-x_n) \Big\langle T \left[ \phi(x_1)\cdots \delta\phi(x_n) \right] \Big\rangle \, ,
\end{align}
where $T$ denotes time-ordering.

\subsection{Single-field consistency relation}

Now let us apply the above general arguments to the Green's functions of the curvature perturbation in the comoving gauge, $H_L|_\text{comoving} \equiv \calR$ under dilatation. Then we briefly recall the simplest consistency relation, i.e. that between the two- and three-point correlation functions of the curvature perturbation. As we have seen in the previous section, dilatation is an exact symmetry so $\Delta = 0$. Further, the correlation functions are evaluated at an equal time $x_1^0 = x_2^0 = \cdots x_n^0 \equiv t$ with respect to the interaction vacuum $|\Omega\rangle$. Thus, integrating \eqref{eq:ward-general} over $y^\mu$ does not make the left-hand side vanishing but gives 
\begin{equation}
\label{eq:ST-R}
\left\langle \left[ Q, \calR(\mathbi{x}_1) \cdots \calR(\mathbi{x}_n) \right] \right\rangle 
= -i \left\langle \delta\calR(\mathbi{x}_1) \calR(\mathbi{x}_2) \cdots \calR(\mathbi{x}_n) \right\rangle - \cdots 
- i \left\langle \calR(\mathbi{x}_1) \cdots \delta\calR(\mathbi{x}_n) \right\rangle \, ,
\end{equation}
where $Q \equiv \int d^3x j^0(t,\mathbi{x})$ is the Noether charge associated with dilatation. Notice that we have suppressed the common time dependence.

As we can see from \eqref{eq:HL-transform}, under dilatation $\delta\calR = -1-x^i\partial_i\calR$. Then, by making use of the translational invariance and moving to the Fourier space, \eqref{eq:ST-R} becomes
\begin{align}
\label{eq:ST-R2}
-i \left\langle \left[ Q, \calR(\mathbi{k}_1) \cdots \calR(\mathbi{k}_n) \right] \right\rangle 
& = -\left[ 3(n-1) + \sum_{i=2}^n \mathbi{k}_i \cdot \nabla_{\mathbi{k}_i} \right] 
\left\langle \calR(\mathbi{k}_1) \cdots \calR(\mathbi{k}_n) \right\rangle 
\nonumber\\
& \quad + \delta^{(3)}(\mathbi{k}_1) \left\langle \calR(\mathbi{k}_2) \cdots \calR(\mathbi{k}_n) \right\rangle 
+ \cdots + \delta^{(3)}(\mathbi{k}_n) \left\langle \calR(\mathbi{k}_1) \cdots \calR(\mathbi{k}_{n-1}) \right\rangle \, ,
\end{align}
with $\mathbi{k}_1 = -\sum_{i=2}^n\mathbi{k}_i$. Here, $\left\langle \calR(\mathbi{k}_1) \cdots \calR(\mathbi{k}_n) \right\rangle$ is the $n$-point correlation function of $\calR$ in the Fourier space, e.g. power spectrum $P_\calR(k)$ for $n=2$ and bispectrum $B_\calR(k_1,k_2,k_3)$ for $n=3$. As we will see, since the dilatation generator $Q$ carries zero momentum and creates $\calR$ from the vacuum, the Slavnov-Taylor identity \eqref{eq:ST-R2} corresponds to soft theorem, relating $(n+1)$- and $n$-point correlation functions of $\calR$ in the factorized form, as implied in the first term of \eqref{eq:ST-R2}. In addition, the Goldstone boson nature of $\calR$ -- shift under dilatation -- results in the $(n-1)$-point function contributions, where one of $n$ $\calR$'s is annihilated into vacuum with zero momentum by $Q$. As far as the vacuum expectation value of $\calR$ vanishes, this contribution to soft theorem becomes evident for $n \geq 2$\footnote{
Indeed, the second line of \eqref{eq:ST-R2} is known as the ``contact terms'', extending the identities to the off-shell case. For more discussion, see Section \ref{subsec:digression}.
}.

To proceed further for the left-hand side of \eqref{eq:ST-R2}, we insert the identity composed of a complete set of orthogonal states denoted by $|m,\mathbi{q}\rangle$, with $m$ being an abstract index denoting independent states collectively, in such a way that~\cite{Assassi:2012zq}
\begin{align}
\label{eq:spectral}
-i \left\langle [Q, \calR(\mathbi{k}_1) \cdots \calR(\mathbi{k}_n)] \right \rangle 
= -i \sum_m \int \frac{d^3q}{(2\pi)^3} \Big[ & \left\langle \Omega \left| Q \right| m,\mathbi{q} \right\rangle 
\left\langle m,\mathbi{q} \left| \calR(\mathbi{k}_1) \cdots \calR(\mathbi{k}_n) \right| \Omega \right\rangle 
\nonumber\\
& - \left\langle \Omega \left| \calR(\mathbi{k}_1) \cdots \calR(\mathbi{k}_n) \right| m,\mathbi{q} \right\rangle 
\left\langle m,\mathbi{q} \left| Q \right| \Omega \right\rangle \Big] \, ,
\end{align}
where for the moment we have made the notation for the vacuum state $|\Omega\rangle$ explicit. If one of the states corresponds to the excitation of the curvature perturbation by applying the operator $\calR(\mathbi{k})$ to the interacting vacuum $|\Omega\rangle$, i.e. 
\begin{equation}
\label{eq:1Rstate}
|m,\mathbi{k}\rangle \supset |\calR,\mathbi{k}\rangle \equiv \frac{1}{P_\calR^{1/2}(k)} \calR(\mathbi{k})|\Omega\rangle \, ,
\end{equation}
where $P_\calR^{-1/2}(k)$ is the normalization factor to satisfy completeness condition $\langle\calR,\mathbi{q}|\calR,\mathbi{k}\rangle = (2\pi)^3\delta^{(3)}(\mathbi{k}-\mathbi{q})$, the Goldstone boson nature of the curvature perturbation becomes evident. Since the charge $Q$ is the generator of the transformation of $\calR$ under dilatation $i[Q,\calR] = \delta\calR$, we obtain
\begin{equation}
\label{eq:goldstone-nature}
\langle \Omega | Q \calR(\mathbi{k}) |\Omega\rangle = \frac{i}{2}(2\pi)^3\delta^{(3)}(\mathbi{k}) + \text{real part} \, .
\end{equation}
This $i/2$ term comes from the Goldstone nature of $\calR$: the {\em shift} under dilatation. Due to the commutator in \eqref{eq:spectral}, the real part that reflects the scalar nature of $\calR$ does not play any role. Also note that the dilatation generator $Q=Q(t)$ only picks $\mathbi{k}=0$, i.e. zero mode. This is in fact not surprising because $Q$ is a function of time only, so spatial dependence should not matter for any commutators involving $Q$. Then, \eqref{eq:spectral} becomes
\begin{equation}
\label{eq:ST-R3}
-i \left\langle [Q, \calR(\mathbi{k}_1) \cdots \calR(\mathbi{k}_n)] \right \rangle = \lim_{q\to0} \frac{\left\langle \calR(\mathbi{q})\calR(\mathbi{k}_1) \cdots \calR(\mathbi{k}_n) \right\rangle}{P_\calR(q)} \, ,
\end{equation}
where we have suppressed the contributions from other states (see however Section~\ref{sec:connectedG-beyond}). Equating \eqref{eq:ST-R2} and \eqref{eq:ST-R3} gives the relations between the connected Green's functions of the curvature perturbation.

The lowest-order connected Green's function corresponds to $n=2$, i.e. the power spectrum $P_\calR(k)$. Using the scale dependence $k^3P_\calR(k) \propto k^{n_\calR-1}$, we find, with $|\mathbi{k}_1| \approx |-\mathbi{k}_2| \equiv k$,
\begin{equation}
\label{eq:consistency_2-3}
\lim_{q\to0} \frac{B_\calR(k_1,k_2,q)}{P_\calR(q)} = (1-n_\calR)P_\calR(k) \, .
\end{equation}
This is the well-known consistency relation between the power spectrum and the squeezed bispectrum of the curvature perturbation~\cite{Maldacena:2002vr,Creminelli:2004yq}.

\subsection{Multi-field consistency relation}

\subsubsection{Consistency relations for the curvature perturbation}
\label{subsubsec:consist}

Having reminded of the consistency relation between the power spectrum and the squeezed bispectrum of the curvature perturbation in single-field inflation that is derived from the Ward identities associated with the dilatation isometry, now we consider how dilatation reveals itself in multi-field inflation. We consider a $n$-dimensional multi-field system in which the field fluctuation $\varphi^a$ can be decomposed into the fluctuation $\pi = -\calR/H$ along the time evolution of the classical solution $\phi_0^a$ and its orthogonal component $\varphi_\perp^a$ as
\begin{equation}
\label{eq:field-decomposition}
\varphi^a(t,\mathbi{x}) = \dot\phi_0^a(t)\pi(t,\mathbi{x}) + \varphi^a_\perp(t,\mathbi{x}) \, ,
\end{equation}
with the orthogonality condition $\dot\phi_{0a}\varphi^a_\perp = 0$. Thus the system contains $n+2$ physical degrees of freedom: the curvature perturbation $\calR$, $n-1$ orthogonal field fluctuations $\varphi_\perp^a$ which satisfy the orthogonality condition $\dot\phi_{0a}\varphi_\perp^a=0$, and two polarizations of the tensor perturbations $\gamma_+$ and $\gamma_\times$. 
We stress that $\varphi_\perp^a$ by construction remains always orthogonal to the curvature perturbation $\calR$, and thus can be interpreted as an independent isocurvature perturbation that can be constrained by observations.
Since at quadratic order the tensor perturbations are decoupled from the scalar ones, any state which contains their excitations defined similar to \eqref{eq:1Rstate} is orthogonal to other degrees of freedom. Meanwhile in general there are quadratic mixing terms between $\calR$ and $\varphi_\perp^a$ and between $\varphi_\perp^a$'s. They provide the leading contributions to the Green's functions $\langle \calR\varphi_\perp^a \rangle$ and $\langle \varphi_\perp^a\varphi_\perp^b \rangle$, so the naive one-particle excitation state $|\varphi_\perp^a,\mathbi{k}\rangle$ defined \`a la \eqref{eq:1Rstate} is not orthogonal to either $|\calR,\mathbi{k}\rangle$ or $|\varphi_\perp^b,\mathbi{k}\rangle$ for $b \neq a$. Instead, we define an $n-1$ orthogonal set of basis
\begin{align}
\label{eq:orthogonalstate}
\left| \widehat{\varphi}_\perp^a,\mathbi{k} \right\rangle 
& \equiv \left[ 1 - \frac{P_{aR}^2(k)}{P_\calR(k)P_a(k)} - \sum_{b \neq a} \frac{P_{ab}^2(k)}{P_a(k)P_b(k)} \right]^{-1/2} 
\nonumber\\
& \quad \times \left[ \left| \varphi_\perp^a,\mathbi{k} \right\rangle 
- \int \frac{d^3q}{(2\pi)^3} \left|\calR,\mathbi{q}\right\rangle 
\left\langle \calR,\mathbi{q} | \varphi_\perp^a,\mathbi{k} \right\rangle 
- \sum_{b \neq a} \int \frac{d^3q}{(2\pi)^3} \left|\widehat{\varphi}_\perp^b,\mathbi{q}\right\rangle 
\left\langle \widehat{\varphi}_\perp^b,\mathbi{q} | \varphi_\perp^a,\mathbi{k} \right\rangle \right] \, ,
\end{align}
where we have assumed that the cross-coupling between $\varphi_\perp^a$ and $\varphi_\perp^b$ for $a \neq b$ [for example, the off-diagonal elements of $M_{ab}^2$: see \eqref{eq:Mab}] is small and can be treated perturbatively, and have presented only leading order terms. This is a reasonable assumption because during multi-field inflation all the dynamical degrees of freedom are light, which means the diagonal components of the mass matrix $M_{aa}^2$ are small and typically the off-diagonal ones $M_{ab}^2$ are even further suppressed. 
We note though that this is not always true with e.g. strong geometric effects~\cite{Gong:2011uw, Renaux-Petel:2015mga} and sharply turning inflationary trajectory~\cite{sharptraj}.
  $P_{a\calR}(k)$ is the cross-correlation function between $\varphi_\perp^a$ and $\calR$, and $P_{ab}(k)$ is that between $\varphi_\perp^a$ and $\varphi_\perp^b$ with $a\neq b$. Notice that writing \eqref{eq:orthogonalstate} does not require any detail of the model: the very existence of the quadratic interaction vertices between physical degrees of freedom demands the cross-correlation functions appear in a specific manner as in \eqref{eq:orthogonalstate}.

Now using the orthogonal one-particle state \eqref{eq:orthogonalstate} for \eqref{eq:spectral} calls for an observation on a nature of $\varphi_\perp^a(t,\mathbi{x})$. Since it does not possess the Goldstone nature under dilatation but is a scalar, there is no shift under dilatation but $\varphi_\perp^a$ transforms as
\begin{equation}
i[Q,\varphi_\perp^a] = \delta\varphi_\perp^a 
= -\mathbi{x} \cdot \nabla \varphi_\perp^a \, .
\end{equation}
The same transformation resulting from the non-Goldstone nature under dilatation holds for the tensor perturbations $\gamma_\lambda$ with $\lambda$ being the polarization index. This gives
\begin{equation}
\label{eq:nonGoldstone}
\langle [Q,\phi] \rangle = 2\Im\langle Q\phi \rangle = 0 
\quad \text{for} \quad 
\phi = \{\varphi_\perp^a,\gamma_\lambda \} \, .
\end{equation}
Thus the contributions of the one-particle state of the tensor perturbations $|\gamma_\lambda,\mathbi{k}\rangle$ and that of the orthogonal field fluctuations $|\varphi_\perp^a,\mathbi{k}\rangle$ to \eqref{eq:spectral} vanish identically. Thus, for the orthogonal state \eqref{eq:orthogonalstate}, only the projection onto $|\calR,\mathbi{q}\rangle$ survives and we find
\begin{align}
\label{eq:multi-consistency}
& -i \int \frac{d^3q}{(2\pi)^3} \Big[ 
\left\langle \Omega \left| Q \right| \widehat\varphi_\perp^a,\mathbi{q} \right\rangle 
\left\langle \widehat\varphi_\perp^a,\mathbi{q} \left| \calR(\mathbi{k}_1) \cdots
\calR(\mathbi{k}_n) \right| \Omega \right\rangle 
- \left\langle \Omega \left| \calR(\mathbi{k}_1) \cdots 
\calR(\mathbi{k}_n) \right| \widehat\varphi_\perp^a,\mathbi{q} \right\rangle 
\left\langle \widehat\varphi_\perp^a,\mathbi{q} \left| Q \right| \Omega \right\rangle \Big]
\nonumber\\
& = \lim_{q\to0} \left[ P_a(q) - \frac{P_{a\calR}^2(q)}{P_\calR(q)} 
- \sum_{b\neq a} \frac{P_{ab}^2(q)}{P_b(q)} \right]^{-1} \frac{-P_{a\calR}(q)}{P_\calR(q)} 
\nonumber\\
& \quad \times \left[ \left\langle \varphi_\perp^a(\mathbi{q})\calR(\mathbi{k}_1) \cdots 
\calR(\mathbi{k}_n) \right\rangle 
- \frac{P_{a\calR}(q)}{P_\calR(q)} \left\langle \calR(\mathbi{q})\calR(\mathbi{k}_1) \cdots
\calR(\mathbi{k}_n) \right\rangle 
- \sum_{b\neq a} \frac{P_{ab}(q)}{P_b(q)} 
\left\langle \varphi_\perp^b(\mathbi{q})\calR(\mathbi{k}_1) \cdots
\calR(\mathbi{k}_n) \right\rangle \right] \, .
\end{align}
This is the contribution of the additional degrees of freedom in multi-field inflation to \eqref{eq:ST-R3}. 
Notice that we have not resorted to any detail of the matter sector in deriving (3.14), so it is true for generic multi-field inflation model.
Especially, for $n=2$, the single-field consistency relation \eqref{eq:consistency_2-3} is modified as
\begin{align}
\label{eq:multiRRR2-3}
\lim_{q\to0} \frac{B_\calR(k_1,k_2,q)}{P_\calR(q)} & = (1-n_\calR)P_\calR(k) 
+ \lim_{q\to0} \left[ P_a(q) - \frac{P_{a\calR}^2(q)}{P_\calR(q)} - \sum_{b\neq a} \frac{P_{ab}^2(q)}{P_b(q)} \right]^{-1} \frac{P_{a\calR}(q)}{P_\calR(q)} 
\nonumber\\
& \quad \times \left[ B_{\calR\calR{a}}(k_1,k_2,q) - \frac{P_{a\calR}(q)}{P_\calR(q)}B_\calR(k_1,k_2,q) - \sum_{b\neq a} \frac{P_{ab}(q)}{P_b(q)} B_{\calR\calR{b}}(k_1,k_2,q) \right] \, .
\end{align}

\subsubsection{Example}

For definiteness, we consider the following explicit action with $n$ scalar fields in the Einstein gravity:
\begin{equation}
S = \int d^4x \sqrt{-g} \left[ \frac{\mpl^2}{2}R - \frac{1}{2}g^{\mu\nu}\partial_\mu\phi^a\partial_\nu\phi_a - V(\phi) \right] \, .
\end{equation}
Then, under the decomposition \eqref{eq:field-decomposition}, in the comoving gauge the quadratic action is given by~\cite{multi-Rint,Gong:2011uw,massmatrix}
\begin{align}
\label{eq:S2}
S_2 = \int d^4x \frac{a^3}{2} \bigg\{ & \dot\varphi_\perp^a\dot\varphi_{\perp a} - \frac{1}{a^2}\partial^i\varphi_\perp^a\partial_i\varphi_{\perp a} - M_{ab}^2\varphi_\perp^a\varphi_\perp^b + 2\epsilon\mpl^2 \left[ \dot\calR^2 - \frac{(\nabla\calR)^2}{a^2} \right] - \frac{4}{H}V_a\varphi_\perp^a\dot\calR 
\nonumber\\
& + \frac{\mpl^2}{4} \left[ \dot\gamma_{ij}^2 - \frac{(\nabla\gamma_{ij})^2}{a^2} \right] \bigg\} \, ,
\end{align}
where $M_{ab}^2$ is
\begin{equation}
\label{eq:Mab}
M_{ab}^2 = V_{ab} - \mathbb{R}_{acdb}\dot\phi_0^c\dot\phi_0^d + (3-\epsilon)\frac{\dot\phi_{0a}}{\mpl}\frac{\dot\phi_{0b}}{\mpl} + \frac{1}{\mpl^2H} \left( \dot\phi_{0a}V_b + \dot\phi_{0b}V_a \right) \, ,
\end{equation}
with $\mathbb{R}_{abcd}$ being the Riemann curvature tensor of the field space. Then the cross-correlation functions can be calculated by treating the mixing terms in \eqref{eq:S2} as perturbations. We use the leading free solutions of the mode functions
\begin{align}
\widehat{\calR}(k,\tau) & = \frac{iH}{\sqrt{4\epsilon k^3}\mpl} (1+ik\tau) e^{-ik\tau} \, ,
\\
\varphi_\perp^a(k,\tau) & = -ie^{i(\nu+1/2)\pi/2} \frac{\sqrt{\pi}}{2} H(-\tau)^{3/2} H_\nu^{(1)}(-k\tau) 
\quad \text{with} \quad 
\nu^2 = \frac{9}{4} - \frac{M_{aa}^2}{H^2} \, ,
\end{align}
where $d\tau=dt/a$ is the conformal time and $H_{\nu}^{(1)}(z)$ is the Hankel function of first kind. Then 
using the in-in formalism~\cite{Weinberg:2005vy}
we can straightly calculate the cross-correlation functions as
\begin{align}
\label{eq:PaR}
P_{a\calR}(k) & = \lim_{-k\tau\to0} \frac{\sqrt{2\pi}}{4} 2^{\nu-3/2} \frac{\Gamma(\nu)}{\Gamma(3/2)} (-k\tau)^{3/2-\nu} \frac{1}{\epsilon\mpl^2k^3} \Re \left[ \int_0^\infty dx V_a x^{-1/2}H_\nu^{(1)}(x)e^{ix} \right] \, ,
\\
\label{eq:Pab}
P_{ab}(k) & = \lim_{-k\tau\to0} \frac{\pi}{8} 2^{\nu_a+\nu_b-3} \frac{\Gamma(\nu_a)\Gamma(\nu_b)}{[\Gamma(3/2)]^2} (-k\tau)^{3-\nu_a-\nu_b} \frac{H}{k^4} \Re \left[ -i \int_0^\infty dx M_{ab}^2 H_{\nu_a}^{(1)}(x) H_{\nu_b}^{(1)}(x) \right] \, .
\end{align}
Note that these cross-correlation functions are the final one evaluated at the end of inflation, as can be noted from the integration range.
In fact, in the $\delta N$ formalism, it was argued~\cite{Tanaka:2010km} that the non-Gaussianity is suppressed both in single- and multi-field inflation. While the $\delta N$ formalism only captures the effects during super-horizon evolution, it is rather a subtle issue how to distinguish sharply sub-, near- and super-horizon contributions in the interactions given in \eqref{eq:S2}.

Beyond the leading order in $M_{ab}^2$ and $V_a/H$, all the correlation functions are corrected more. For example, the correction to ${\cal P}_\calR \equiv k^3P_\calR/(2\pi^2)$ is of ${\cal O}\big[(V_a/H)^2\big]$ and is given by~\cite{Pcorrections}
\begin{align}
\Delta{\cal P}_\calR = \frac{\pi}{8\epsilon^2}\frac{1}{H^2\mpl^4}\frac{1}{k^3}
& \left\{ \left|\int_0^\infty dx V_a x^{-1/2} H_\nu^{(1)}(x)e^{ix} \right|^2 \right. 
\nonumber\\
& \left. - 2\Re \left[ \int_0^\infty dx_1 V_a e^{-ix_1}x_1^{-1/2}H_\nu^{(1)}(x_1)
\int_{x_1}^\infty dx_2 V_a e^{-ix_2}x_2^{-1/2}H_\nu^{(2)}(x_2) \right] \right\} \, .
\end{align}
However, we expect that the consistency relation determined from the symmetry principle in the form of \eqref{eq:multiRRR2-3} still holds under the same order of corrections for each of two- and three-point functions.

Now let us further consider the simplest two-field case. While the quadratic action for the curvature perturbation and the tensor perturbations remain the same, in \eqref{eq:S2} the orthogonal field $\varphi$ contributes, including the mixing with $\calR$~\cite{multi-Rint}, 
\begin{equation}
S_2 \supset \int d^4x \frac{a^3}{2} \left[ \dot\varphi^2 - \frac{(\nabla\varphi)^2}{a^2} - M^2\varphi^2 + 4\dot\theta\frac{\dot\phi_0}{H} \varphi\dot\calR \right] \, ,
\end{equation}
where $\dot\theta$ is the angular velocity of the trajectory defined by
\begin{equation}
\dot\theta \equiv -\frac{V_N}{\dot\phi_0} \, ,
\end{equation}
with $V_N$ denoting the projection of the potential derivative onto the orthogonal direction to the background trajectory. The mass scale $M^2$ is given by
\begin{equation}
M^2 = V_{NN} + \epsilon\mpl^2\mathbb{R} - \dot\theta^2 \, .
\end{equation}
In this case, the only cross-correlation function is that between $\calR$ and $\varphi$ and is given by
\begin{equation}
P_{\varphi\calR} = \lim_{-k\tau\to0} \frac{\sqrt{\pi}}{2} 2^{\nu-3/2} 
\frac{\Gamma(\nu)}{\Gamma(3/2)} (-k\tau)^{3/2-\nu} \frac{H}{\sqrt{\epsilon}\mpl k^3} 
\Re \left[ - \int_0^\infty dx \dot\theta x^{-1/2}H_\nu^{(1)}(x)e^{ix} \right] \, ,
\end{equation}
and \eqref{eq:multiRRR2-3} is simplified to
\begin{equation}
(1-n_\calR)P_\calR(k) = \lim_{q\to0} 
\left[ P_\varphi(q) - \frac{P_{\varphi\calR}^2(q)}{P_\calR(q)} \right]^{-1} 
\left[ \frac{P_{\varphi}(q)}{P_\calR(q)} B_\calR(k_1,k_2,q)
- \frac{P_{\varphi\calR}(q)}{P_\calR(q)} B_{\calR\calR{\varphi}}(k_1,k_2,q) \right] \, .
\end{equation}
We note here that in the decoupling limit where the angular velocity $\dot{\theta}$ is negligibly small, the single-field consistency relation \eqref{eq:consistency_2-3} is recovered.

\subsubsection{Beyond the consistency relations for the curvature perturbation}
\label{sec:connectedG-beyond}

What \eqref{eq:spectral} tells us is that, even if we only consider the connected $n$-point Green's function of the curvature perturbation $\calR$, its Ward identities receive contributions from independent states of the system other than $|\calR,\mathbi{k}\rangle$ in a specific manner. These independent states include not only independent degrees of freedom, but also different ``excitations'' of the same degree of freedom. As we have seen, the shift transformation of $\calR$ under dilatation invites us to interpret $\calR$ as a Goldstone boson associated with dilatation, or the fluctuation along the direction of transformation of the vacuum generated by $Q$. As a result, $Q$ does not annihilate the vacuum but creates the excitation of $\calR$, as given by \eqref{eq:1Rstate}. On the other hand, the non-Goldstone nature under dilatation \eqref{eq:nonGoldstone} of the orthogonal field fluctuations $\varphi_\perp^a$ and the tensor perturbations $\gamma_\lambda$ prohibits their contributions to the consistency relations. Only the projection of $\varphi_\perp^a$ onto $\calR$, non-zero due to the quadratic interaction in \eqref{eq:S2}, survives which is the only additional contribution among the one-particle states of other degrees of freedom.

Now, in principle, the dilatation generator $Q$ can create any number of $\calR$ from the vacuum, provided that total spatial momentum vanishes. Typically, such kind of matrix element of operator contributing to multi-particle production  from the vacuum is suppressed in different contexts. For example, in the presence of perturbatively small interaction, probability for multi-particle production from the vacuum induced by quantum field operator is just given by perturbatively small decay rate of quantum state created by the corresponding operator. In the case of QCD, where strong interaction prevents us from perturbative prediction using quark-gluon interactions, the large $N$ expansion with $N$ being the number of colors provides a qualitative justification that spontaneously broken chiral charge produces one pion dominantly~\cite{Witten:1979kh}. In our case, the existence of  $\calR$ is parametrized by non-zero slow-roll parameters representing the deviation from dS space-time. If the space-time background is given by exact dS, the scale invariance is restored, and the ``Goldstone boson'' $\calR$ does not exist. As our universe is almost scale invariant, small slow-roll parameters play the role of the expansion parameter for cosmological perturbations. To leading slow-roll order, the matrix element of $Q$ is dominated by the creation of single $\calR$ from the vacuum: the contributions of multi-particle states are slow-roll suppressed.

For example, the contribution of the ``two-particle'' state of the curvature perturbation $\left|\calR^2,\mathbi{k}\right\rangle$ to \eqref{eq:spectral} is
\begin{align}
\label{eq:R2state-result}
& -i \int \frac{d^3q}{(2\pi)^3} 
\Big[ \left\langle\Omega\left| Q \right| \calR^2,\mathbi{q} \right\rangle 
\left\langle \calR^2,\mathbi{q} \left| \calR(\mathbi{k}_1) \cdots \calR(\mathbi{k}_n) \right|\Omega\right\rangle 
- \left\langle \Omega \left| \calR(\mathbi{k}_1) \cdots \calR(\mathbi{k}_n) \right|\calR^2,\mathbi{q} \right\rangle 
\left\langle \calR^2,\mathbi{q} \left| Q \right| \Omega \right\rangle \Big]
\nonumber\\
& = \lim_{\mathbi{q}\to0} \frac{1}{2\tau_{\rm NL}P_\calR(q)} \left[ \int \frac{d^3p}{(2\pi)^3} P_\calR(p) \right]^{-2}
\int \frac{d^3p}{(2\pi)^3}  
\langle \calR(\mathbi{k}_1) \cdots \calR(\mathbi{k}_n) \calR(\mathbi{p})\calR(\mathbi{q}-\mathbi{p}) \rangle 
(1-n_\calR) P_\calR(|\mathbi{q}-\mathbi{p}|) \, ,
\end{align}
where the non-linear parameter $\tau_{\rm NL}$, assumed to be constant, is given in terms of the collapsed limit of the trispectrum~\cite{Byrnes:2006vq}:
\begin{equation}
\tau_{\rm NL} \equiv \frac{1}{4} \lim_{|\mathbi{k}_1+\mathbi{k}_2|\to0} 
\frac{T_\calR(\mathbi{k}_1,\mathbi{k}_2,\mathbi{k}_3,\mathbi{k}_4)}{P_\calR(k_1)P_\calR(k_3)P_\calR(|\mathbi{k}_1+\mathbi{k}_2|)} \, .
\end{equation}
We present the detailed steps to derive \eqref{eq:R2state-result} in Appendix~\ref{app:2particle}. Comparing with the result from the one-particle state $|\calR,\mathbi{q}\rangle$ given by \eqref{eq:ST-R3}, applying the two-particle state gives the result which
\begin{itemize}
\item involves the next higher-order connected Green's function, viz. $(n+2)$-point correlation function, and
\item involves the spectral index of the power spectrum, which is $\calO(\epsilon)$.
\end{itemize}
Thus the contribution of the two-particle state \eqref{eq:R2state-result} is, compared to the one-particle state result \eqref{eq:ST-R3}, slow-roll suppressed and is sub-dominant. We can proceed further similarly for the three- and more-particle states $\left| \calR^n,\mathbi{k} \right\rangle$ with $n\geq3$, whose contributions are even further slow-roll suppressed.

We can proceed similarly for the states that contain mixed excitations. For example consider the operator product in the configuration space of the form $\calR(\mathbi{x})\phi(\mathbi{y})$ with $\phi = \{\varphi_\perp^a,\gamma_\lambda\}$. This corresponds in the momentum space to the state that contains the excitations of one $\calR$ and one $\phi$ particles. Then
\begin{align}
i[Q,\calR(\mathbi{x})\phi(\mathbi{y})] & = i[Q,\calR(\mathbi{x})]\phi(\mathbi{y}) + i\calR(\mathbi{x})[Q,\phi(\mathbi{y})]
\nonumber\\
& = \left[ -i - x^i \frac{\partial\calR(\mathbi{x})}{\partial x^i} \right] \phi(\mathbi{y}) - \calR(\mathbi{x}) y^i \frac{\partial\phi(\mathbi{y})}{\partial y^i} \, ,
\end{align} 
so the expectation value of this commutator is
\begin{equation}
\Big\langle [Q,\calR(\mathbi{x})\phi(\mathbi{y})] \Big\rangle = ix^i \left\langle \frac{\partial\calR(\mathbi{x})}{\partial x^i}\phi(\mathbi{y}) \right\rangle + i y^i \left\langle \calR(\mathbi{x})\frac{\partial\phi(\mathbi{y})}{\partial y^i} \right\rangle \, ,
\end{equation}
which is non-zero only for $\phi = \varphi_\perp^a$ since the tensor perturbations are decoupled, and is proportional to the product of the spectral index of the cross-correlation power spectrum $P_{a\calR}(k)$ and the power spectrum itself. Given that in multi-field inflation the orthogonal fields are also slowly varying, the contribution from this state is in general further suppressed compared to that from $|\calR,\mathbi{k}\rangle$.

Another observation we can make regarding the dilatation isometry is that any mixed $n$-point correlation functions including $\varphi_\perp^a$ and $\gamma_\lambda$ are not affected. For example consider the bispectrum $\langle\calR(\mathbi{q})\calR(\mathbi{k}_1)\varphi_\perp^a(\mathbi{k}_2)\rangle$, assuming the same form of the spectral index of the cross-correlation power spectrum $k^3P_{a\calR}(k) \propto k^{n_{a\calR}-1}$. Then \eqref{eq:spectral} read
\begin{align}
\lim_{q\to0} \frac{B_{\calR \calR a}(q,k_1,k_2)}{P_\calR(q)} & = (1-n_{a\calR})P_{a\calR}(k) 
+ \lim_{q\to0} \left[ P_b(q) - \frac{P_{b\calR}^2(q)}{P_\calR(q)} - \sum_{c\neq b} \frac{P_{bc}^2(q)}{P_c(q)} \right]^{-1} \frac{P_{b\calR}(q)}{P_\calR(q)} 
\nonumber\\
& \quad \times \left[ B_{b\calR{a}}(q,k_1,k_2) - \frac{P_{b\calR}(q)}{P_\calR(q)}B_{\calR\calR a}(q,k_1,k_2) - \sum_{c\neq b} \frac{P_{bc}(q)}{P_c(q)} B_{c\calR{a}}(q,k_1,k_2) \right] \, .
\end{align}
Here, the Goldstone boson nature of the curvature perturbation $\calR$ comes in through the complete set of orthogonal states as the second line of \eqref{eq:ST-R2} is absent. Hence, the form of the consistency relation is the same as that of the curvature perturbation.

\section{1PI Green's function point of view}
\label{Sec:1pi}
\setcounter{equation}{0}

Symmetries of the physical system are encoded in the interaction vertices in the Lagrangian. These vertices receive quantum correction from loop effects.
Among the connected Green's functions, we define 1PI Green's function as a subset such that when represented in terms of the Feynman diagram, any single cut of an internal line does not spoil the connectedness. Then, a connected Green's function is regarded as a tree diagram when 1PI diagrams shrink to a point. Indeed, taking $\hbar \to 0$ limit, 1PI Green's functions are reduced to interaction vertices in the bare Lagrangian. In this sense, 1PI Green's functions correspond to quantum-corrected interaction vertices in the connected Green's function, hence the basic units encoding symmetry. Therefore, it is meaningful to obtain the 1PI version of the Slavnov-Taylor identity -- the relations between 1PI Green's functions under the dilatation symmetry, which should be consistent with the results obtained in Section~\ref{Sec:cgf}.

\subsection{Quantum effective action and consistency relations}
\label{subsec:1PIconsist}

The 1PI Green's functions are generated by the quantum effective action, which is obtained as follows. After gauge fixing, and integrating out auxiliary fields, the functional integral representation of the vacuum-to-vacuum amplitude is given by the generating functional $Z[J]$:
\begin{align}
\label{eq:genfunc}
Z[T^{ij},J_a,\bar\chi,\chi] & =
\int \calD{h}_{ij} \calD \widehat{\varphi}_\perp^a \calD\eta \calD\bar\eta
\exp \left\{ i \left[ S \left[ h_{ij},\widehat{\varphi}_\perp^a,\eta,\bar\eta \right]
+ \int d^4x \big( h_{ij}T^{ij} + \widehat{\varphi}_\perp^aJ_a + \bar\chi\eta + \chi\bar\eta \big)
\right] \right\}
\nonumber\\
& \equiv e^{iG_c[T^{ij},J_a,\bar\chi,\chi]} \, ,
\end{align}
where $G_c$ generates connected Green's functions. The sources $J_a$ for scalar fields are attached to $\widehat{\varphi}_\perp^a$ rather than $\varphi_\perp^a$, since $\widehat{\varphi}_\perp^a$ are responsible for creation and annihilation of independent orthogonal states, as shown in Section \ref{subsubsec:consist}. As we have seen in Section~\ref{sec:gauge-iso}, under dilatation the gauge conditions remain invariant, so  the ghost terms can be decoupled (see however Section~\ref{subsec:digression}). Thus we can ignore the ghost fields $\eta$ and $\bar\eta$ as well as the external currents associated with them, $\bar\chi$ and $\chi$. The classical fields are defined by
\begin{equation}
h_{ij}(x, T^{ij}, J_a) = \frac{\delta}{\delta T^{ij}}G_c [T^{ij}, J_a]
\quad \text{and} \quad
\widehat{\varphi}^a_\bot(x, T^{ij}, J_a)=\frac{\delta}{\delta J_a} G_c [T^{ij}, J_a] \, ,
\end{equation}
from which the quantum effective action $\Gamma$ is obtained through the Legendre transformation~\cite{Weinberg1995}:
\begin{equation}
\Gamma[h_{ij},\widehat{\varphi}^a_\bot] \equiv G_c[T^{ij}, J_a]
- \int d^4 x \big( T^{ij} h_{ij} + \widehat{\varphi}_\bot^a J_a \big) 
\equiv G_c[T^{ij},J_a] - S_\text{ext} \, .
\end{equation}
Then $\Gamma$ becomes the generating functional for the 1PI Green's function: expanding $\Gamma$ in terms of the ``classical fields'' $h_{ij}$ and $\widehat{\varphi}_\bot^a$, the coefficients given by $n$ derivatives with respect to the classical fields at vanishing classical field values correspond to the $n$-point 1PI Green's functions. The external currents $T^{ij}$ and $J_a$, associated with $h_{ij}$ and $\widehat{\varphi}_\perp^a$ respectively, are thus obtained as
\begin{equation}
\frac{\delta \Gamma}{\delta h_{ij}} = -T^{ij}
\quad \text{and} \quad 
\frac{\delta \Gamma}{\delta \widehat{\varphi}_\bot^a} = -J_a \, .
\end{equation}

The Ward identity states that under infinitesimal changes in the fields, the generating functional is invariant:
\begin{equation}
\delta{Z} = \int \calD{h}_{ij}\calD\varphi_\perp^a e^{iS + iS_\text{ext}} i\delta{S}_\text{ext}
= 0 \, ,
\end{equation}
since the action $S$ is already invariant. The changes in the fields, $h_{ij}$ and $\widehat{\varphi}_\perp^a$, arise under the spatial coordinate transformation $x^i \to x^i + \xi^i$ given by \eqref{eq:gaugetransformation}. This gives
\begin{equation}
\delta{Z} = \int d^3x \xi^i \bigg[ 2i\partial_jT^{jk} 
- T^{ij}\partial_k \bigg( \frac{\delta}{\delta{T}^{ij}} \bigg)
+ \partial_j \bigg( T^{ij}\frac{\delta}{\delta{T}^{ik}} \bigg) 
- \partial_j \bigg( T^{ik}\frac{\delta}{\delta{T}^{ij}} \bigg)
- J_a \partial_k \bigg( \frac{\delta}{\delta{J}^a} \bigg) \bigg] e^{iG_c} \, ,
\end{equation}
where we have raised the indices of the current $T^{ij}$ up, since as set in Section~\ref{sec:gauge-iso} we have singled out the scale factor so that upper and lower indices for $h_{ij}$, and accordingly those for $T^{ij}$, do not make any difference. Thus, $\delta{Z} = 0$ gives the following non-trivial identity~\cite{1pi-identities,Berezhiani:2013ewa}:
\begin{align}
\label{eq:mastereq}
-2 \partial_j \bigg( \frac{1}{6}\delta_{jk} \frac{\delta\Gamma}{\delta\calR} 
+ \frac{\delta\Gamma}{\delta\gamma_{jk}} \bigg) 
+ \calR_{,k}\frac{\delta\Gamma}{\delta\calR} + \gamma_{ij,k}\frac{\delta\Gamma}{\delta\gamma_{ij}}
- \partial_j \bigg( \gamma_{ik}\frac{\delta\Gamma}{\delta\gamma_{ij}} \bigg)
+ \gamma_{ij}\partial_j\frac{\delta\Gamma}{\delta\gamma_{ik}} 
+ \partial_k\widehat{\varphi}_\perp^a\frac{\delta\Gamma}{\delta\widehat{\varphi}_\perp^a}  = 0 \, ,
\end{align}
where we have used an identity for a square matrix $\mathbb{M}$ that $\partial{\rm Tr}(\mathbb{M})/\partial\mathbb{M} = 1\!\!1$. Notice that in deriving \eqref{eq:mastereq} we have not made any approximation, thus \eqref{eq:mastereq} is exact for both the curvature perturbation $\calR$ and the tensor perturbations $\gamma_{ij}$. As mentioned above, taking derivatives of \eqref{eq:mastereq} with respect to the classical fields $\calR$, $\gamma_{ij}$ and $\widehat{\varphi}_\perp^a$ gives the 1PI Green's functions.

First let us consider taking two derivatives of \eqref{eq:mastereq} with respect to $\calR$ and moving to the momentum space. We can then check the lowest-order consistency relation. After setting $\calR = \gamma_{ij} = \widehat{\varphi}_\perp^a = 0$, we obtain
\begin{align}
\label{eq:verticesRR}
& \frac{1}{3} q_{i} \Gamma_{\calR\calR\calR}(\mathbi{q},\mathbi{k},-\mathbi{q}-\mathbi{k})
+ 2 q_{j} \Gamma_{\gamma_{ij}\calR\calR}(\mathbi{q},\mathbi{k},-\mathbi{q}-\mathbi{k})
+ k_{i} \Big[ \Gamma_{\calR\calR}(|\mathbi{k}+\mathbi{q}|) - \Gamma_{\calR\calR}(k) \Big]
- q_{i} \Gamma_{\calR\calR}(k) = 0 \, ,
\end{align}
where we have assumed translational invariance of the effective action, and $\Gamma_{\calR\calR\calR}$ denotes the functional derivative of $\Gamma$ with respect to three $\calR$'s evaluated at $\calR = \gamma_{ij} = \widehat{\varphi}_\perp^a = 0$, and similarly for other subscripts like $\Gamma_{\gamma_{ij}\calR\calR}$. Up to this point, we have not imposed any specific configurations for the momenta in \eqref{eq:verticesRR}: this formula holds for any generic $\mathbi{k}$ and $\mathbi{q}$. To check the  consistency relation in the limit, say, $q\to0$, we may expand \eqref{eq:verticesRR} to have
\begin{equation}
\label{eq:verticesRR-2}
\lim_{q\to0}
q_{j} \bigg[ \frac{\delta_{ij}}{3} \Gamma_{\calR\calR\calR}(\mathbi{q},\mathbi{k},-\mathbi{k}) 
+ 2\Gamma_{\gamma_{ij}\calR\calR}(\mathbi{q},\mathbi{k},-\mathbi{k})
+ k_{i} \frac{\partial}{\partial{k_{j}}} \Gamma_{\calR\calR}(k) 
- \delta_{ij}\Gamma_{\calR\calR}(k) \bigg] = 0 \, .
\end{equation}
Since the formula in the square brackets carries two spatial indices, we can decompose -- see however the following subsection -- it into the trace and traceless parts: for an arbitrary $K_{ij}$, we can write the trace part $K$ and the traceless part $\overline{K}_{ij}$ as
\begin{equation}
K \equiv K_{ii} 
\quad \text{and} \quad
\overline{K}_{ij} = K_{ij} - \frac{\delta_{ij}}{3}K .
\end{equation}
Then \eqref{eq:verticesRR-2} is decomposed into 
\begin{align}
\label{eq:sss-consistency}
\text{Trace: } & \lim_{q\to0} \Gamma_{\calR\calR\calR}(\mathbi{q},\mathbi{k},-\mathbi{k}) 
- \Big( 3 - \mathbi{k}\cdot\nabla_{\mathbi{k}} \Big) \Gamma_{\calR\calR}(k) = 0 \, ,
\\
\label{eq:tss-consistency}
\text{Traceless: } & \lim_{q\to0} 2\Gamma_{\gamma_{ij}\calR\calR}(\mathbi{q},\mathbi{k},-\mathbi{k}) 
+ \bigg( k_{i} \frac{\partial}{\partial{k_{j}}} - \frac{\delta_{ij}}{3} \mathbi{k}\cdot\nabla_{\mathbi{k}} \bigg) \Gamma_{\calR\calR}(k) = 0 \, .
\end{align}
In the same manner, we can take derivatives of \eqref{eq:mastereq} with respect to other degrees of freedom or their combinations. For example, taking derivatives with respect to $\calR$ and $\widehat{\varphi}_\perp^a$ and two $\widehat{\varphi}_\perp^a$'s lead to, respectively,
\begin{align}
\label{eq:verticesRa}
& \frac{1}{3} q_{i} \Gamma_{\calR\calR a}(\mathbi{q},\mathbi{k},-\mathbi{q}-\mathbi{k})
+ 2 q_{j} \Gamma_{\gamma_{ij}\calR a}(\mathbi{q},\mathbi{k},-\mathbi{q}-\mathbi{k})
\nonumber\\
& \quad
+ k_{i} \Big[ \Gamma_{\calR a}(|\mathbi{k}+\mathbi{q}|) - \Gamma_{\calR a}(k) \Big]
- q_{i} \Gamma_{\calR\calR}(k) = 0 \, ,
\\
\label{eq:verticeab}
& \frac{1}{3} q_{i} \Gamma_{\calR ab}(\mathbi{q},\mathbi{k},-\mathbi{q}-\mathbi{k})
+ 2 q_{j} \Gamma_{\gamma_{ij}ab}(\mathbi{q},\mathbi{k},-\mathbi{q}-\mathbi{k})
\nonumber\\
& \quad
+ k_{i} \Big[ \Gamma_{ab}(|\mathbi{k}+\mathbi{q}|) - \Gamma_{ab}(k) \Big]
- q_{i} \Gamma_{\calR\calR}(k) = 0 \, ,
\end{align}
from which in the squeezed limit $q\to0$ we find the following trace components:
\begin{align}
\lim_{q\to0} \Gamma_{\calR\calR a}(\mathbi{q},\mathbi{k},-\mathbi{k}) 
- \Big( 3 - \mathbi{k}\cdot\nabla_{\mathbi{k}} \Big) \Gamma_{\calR a}(k) & = 0 \, ,
\\
\lim_{q\to0} \Gamma_{\calR ab}(\mathbi{q},\mathbi{k},-\mathbi{k}) 
- \Big( 3 - \mathbi{k}\cdot\nabla_{\mathbi{k}} \Big) \Gamma_{ab}(k) & = 0 \, ,
\end{align}
and the following traceless components:
\begin{align}
\lim_{q\to0} 2\Gamma_{\gamma_{ij}\calR a}(\mathbi{q},\mathbi{k},-\mathbi{k}) 
+ \bigg( k_{i} \frac{\partial}{\partial{k_{j}}} - \frac{\delta_{ij}}{3} \mathbi{k}\cdot\nabla_{\mathbi{k}} \bigg) \Gamma_{\calR a}(k) & = 0 \, .
\\
\lim_{q\to0} 2\Gamma_{\gamma_{ij}ab}(\mathbi{q},\mathbi{k},-\mathbi{k}) 
+ \bigg( k_{i} \frac{\partial}{\partial{k_{j}}} - \frac{\delta_{ij}}{3} \mathbi{k}\cdot\nabla_{\mathbi{k}} \bigg) \Gamma_{ab}(k) & = 0 \, .
\end{align}
These correspond to the lowest-order consistency relations satisfied by the 1PI Green's function. Of course, physical observables are not 1PI, but connected Green's functions and we need to check whether the approaches using the different Green's functions are equivalent. It can be checked through the fact that the connected Green's functions are constructed by connecting the 1PI Green's functions with two-point connected Green's functions. Details can be found in Appendix \ref{app:equiv}.

\subsection{More lessons from 1PI approach}
\label{subsec:digression}

The consistency relations from the connected Green's function approach studied in Section \ref{Sec:cgf} rely on the vacuum structure of the theory, creating the curvature perturbation with zero momentum under the action of the spontaneously broken dilatation charge. Therefore, the three-point function is related to the two-point function by taking one of three external momenta to be zero, i.e. in the soft limit. On the other hand, in the 1PI approach, we are interested in the symmetries of the Lagrangian rather than the vacuum. It thus gives the 1PI Slavnov-Taylor identities for full diffeomorphim, no matter whether it is spontaneously broken or not -- note that while spontaneous broken dilatation symmetry is reflected in the shift behavior of the curvature perturbation $\calR$, as argued in Section \ref{sec:gauge-iso} not all diffeomorphism are consistent with gauge fixing conditions. For complete description, the variations of the gauge fixing conditions should be taken into account in addition, hence more systematic analysis can be made in terms of the BRST symmetry. Of course, as we have seen, under the comoving gauge we are working with, dilatation does not alter the gauge fixing conditions so the considerations in Section \ref{subsec:1PIconsist} are enough to give the consistency relations that holds even in the quantum level from dilatation. This 1PI consistency relation corresponds to the trace part \eqref{eq:sss-consistency}, because the dilatation transformation $\xi^i = x^i$ automatically picks the trace of \eqref{eq:mastereq} since $\partial_j\xi^i = \delta^i{}_j$. The traceless part \eqref{eq:tss-consistency}, however, corresponds to the consistency relation associated with another spatial coordinate transformation that holds either only at tree level if gauge fixing contributions are not supplemented to \eqref{eq:tss-consistency} \cite{Berezhiani:2013ewa} or at quantum level including all loop corrections if the corresponding coordinate transformation also preserves the comoving gauge conditions.

We also note that the 1PI Slavnov-Taylor identities are not restricted to the soft limit. Indeed,  \eqref{eq:verticesRR} does not impose the soft limit of the momentum $q$ unlike the connected Green's function approach in which the soft limit is automatically imposed since the dilatation charge does not carry any momentum. So we may expect that the 1PI Slavnov-Taylor identities provide other consistency relations, for example for general momentum configurations such as the equilateral one. However, such relations are apt to be contaminated by an arbitrary transverse tensor, because there is an ambiguity in trading the vector relation to the tensor one as done in \eqref{eq:verticesRR-2}. In principle, the tensor in the square bracket of \eqref{eq:verticesRR-2} cannot be naively identified with zero, since it may well contain any tensor $A_{ij}$ satisfying $q_j A_{ij}=0$.  The consistency relation in the soft limit implies $\lim_{q\to0}A_{ij} = 0$, say $A_{ij}\sim {\cal O}(q^2)$ as discussed in \cite{Berezhiani:2013ewa}. Since the connected Green's function approach only provides soft limit relations, such unknown tensor is not restricted by symmetry principles alone but contains model dependence that is not negligible in the non-soft limit. For example, in the equilateral configuration $q=k=|\mathbi{k}+\mathbi{q}|$, we can write \eqref{eq:verticesRR} including the unconstrained tensor $A_{ij}$ as
\begin{align}
0 & = \frac13 q_i \Gamma_{\calR\calR\calR}(\mathbi{q},\mathbi{k},-\mathbi{k}-\mathbi{q})
+ 2 q_j\Gamma_{\gamma_{ij}\calR\calR}(\mathbi{q},\mathbi{k},-\mathbi{k}-\mathbi{q})
- q_i\Gamma_{\calR\calR}(k)
\nonumber\\
& = q_i \left[ \frac{\delta_{ij}}{3}\Gamma_{\calR\calR\calR}(\mathbi{q},\mathbi{k},-\mathbi{k}-\mathbi{q})
+ 2 \Gamma_{\gamma_{ij}\calR\calR}(\mathbi{q},\mathbi{k},-\mathbi{k}-\mathbi{q})
- \delta_{ij}\Gamma_{\calR\calR}(k) + A_{ij} \right] \, .
\end{align}
Taking the trace part of the bracket, we obtain 
\begin{equation}
\label{eq:consistency-equil}
\Gamma_{\calR\calR\calR}(\mathbi{q},\mathbi{k},-\mathbi{k}-\mathbi{q}) = 3\Gamma_{\calR\calR}(k) - A \, ,
\end{equation}
where $A \equiv A_{ii}$. Since $B_\calR=P_\calR(q)P_\calR(k)P_\calR(|\mathbi{k}+\mathbi{q}|)\Gamma_{\calR\calR\calR}+\cdots$ and $P_\calR=-\Gamma_{\calR\calR}^{-1}+\cdots$, we find that the bispectrum in a configuration well away from the squeezed limit, e.g. equilateral one, contains a model-dependent factor $A$ to be specified additionally. For example, in k-inflation type models \cite{k-inflation} whose matter sector is given by
\begin{equation}
S_m = \int d^4x \sqrt{-g} P(X,\phi)
\end{equation}
with $X \equiv -\partial^\mu\phi\partial_\mu\phi/2$, the non-linear parameter in the equilateral configuration is boosted by the speed of sound, $f_{\rm NL} \sim 1/c_s^2$~\cite{Chen:2006nt}, with 
\begin{equation}
c_s^2 \equiv \frac{P_X}{P_X+2XP_{XX}}
\end{equation}
describing the matter sector of the theory.

Another observation is that the consistency relations in the soft limit are independent of detail of the matter sector. In both approaches using the connected and 1PI Green's functions, what is essential is to correctly identify the degrees of freedom which are mutually orthogonal and independent from each other. They are found by subtracting the projection onto the other degrees of freedom due to the quadratic interaction terms. The form of the orthogonal degrees of freedom constructed in that manner -- no matter in the disguise of state \eqref{eq:orthogonalstate} or operator \eqref{eq:orthogonalop} -- is thus universal: given that there are quadratic interactions, the orthogonal degrees of freedom and in turn the consistency relations should be written in such a way that the cross-correlation functions appear only in a specific manner. The detail of the interaction terms is necessary only when we are to find explicitly the cross-correlation functions like \eqref{eq:PaR} and \eqref{eq:Pab}. However  as we discussed above, this model independence is lost when we are away from the soft limit which is possible in the 1PI Green's function approach.

We finally make some comments about different observables in cosmology and particle physics. In particle physics with four-dimensional Minkowski background, all the physical processes are described in terms of the S-matrix -- the transition amplitude from initial to final asymptotic states. The S-matrix is obtained by the Lehmann-Symanzik-Zimmerman (LSZ) formalism~\cite{Lehmann:1954rq}, where on-shell one-particle states are extracted from the Green's functions as external (asymptotic) states. Through this procedure, the contact terms in the Slavnov-Taylor identities for the connected Green's functions disappear, such as the contributions of the $(n-1)$-point correlation functions in \eqref{eq:ST-R2}. In the 1PI approach, four-dimensional translation invariance would induce the identity corresponding to \eqref{eq:mastereq}, with the absence of the curvature perturbation $\calR$. Then the term $-2\partial_j \delta \Gamma/\delta \gamma_{ij}$ corresponds to $-2\partial_\nu \delta \Gamma/\delta g_{\mu\nu}=-2\partial_\nu \big( \mpl^2 G^{\mu\nu}-T_m^{\mu\nu} \big)$, where $T_m^{\mu\nu}$ is the matter energy-momentum tensor, and the other terms vanish on the mass shell, just stating the equivalence between gravitation and geometry. On the other hand, in cosmological perturbation theory, main observables are connected Green's functions, and the procedure like the LSZ formalism that constrains external states on the mass shell does not exist. For this reason, the Slavnov-Taylor identities appear in the off-shell form. Especially, the fact that cosmological perturbation theory does not specify asymptotic on-shell states may give rise to potential subtle issues concerning renormalization. In the unitary gauge of spin-1 gauge theory, massive gauge boson propagator makes the Green's function ultraviolet divergent even if the theory is renormalizable\footnote{
We thank Daniel Chung for reminding us of this issue.
}. This inconvenient renormalization is tamed after taking on-shell limit in the S-matrix \cite{onshell-smatrix}. Since some of diffeomorphism are spontaneously broken during the inflation as well, it is challenging to find out an appropriate description of quantum-corrected cosmological perturbation theory in the comoving gauge, the analogy of the unitary gauge in particle physics.

\section{Conclusions}
\label{Sec:conclusion}

In this article, we have considered the consequences of the spatial coordinate transformation in multi-field inflation. Among the spontaneously broken dS isometries, only dilatation commutes with the comoving gauge conditions \eqref{eq:comovinggauge} while SCTs do not. Thus the Slavnov-Taylor identities associated with dilatation is exact even at quantum level, not necessarily incorporating the BRST symmetry. Other gauge conditions, e.g. the flat gauge in which all the (scalar) degrees of freedom are given to the matter sector, are not invariant under both dilatation and SCTs, so the spatial coordinate reparametrization symmetry cannot be reflected.

We have derived the Slavnov-Taylor identities associated with dilatation and the resulting consistency relations in two ways. In the connected Green's function approach, as can be read from the Slavnov-Taylor identities \eqref{eq:ST-R} [or even more fundamentally \eqref{eq:ward-general}], from the beginning we are restricted to consider the soft limit of the consistency relations because the dilatation charge $Q$ brings the shift of the curvature perturbation \eqref{eq:goldstone-nature} with vanishing momentum. More importantly, the other independent degrees of freedom do not naively participate in the consistency relations, but only the quadratic interactions with the curvature perturbation contribute. This is because only the curvature perturbation possesses the Goldstone nature under dilatation. Thus the cross-correlation functions should appear in a specific manner as \eqref{eq:multi-consistency} [or \eqref{eq:multiRRR2-3} at the lowest order] to extract the orthogonal component, and in that sense the consistency relations in the soft limit are universal and independent of the detail of the model even in multi-field inflation.

Using the 1PI Green's functions which encode the symmetry of the system, i.e. general covariance in the current case, we have obtained an alternative form of the Slavnov-Taylor identities \eqref{eq:mastereq} from which the consistency relations follow. In deriving \eqref{eq:mastereq} we have not made any approximation even for the tensor sector thus \eqref{eq:mastereq} is exact, given that the spatial coordinate transformation under consideration does not change the gauge conditions. The resulting consistency relations like \eqref{eq:verticesRR} hold for generic momentum configurations. By taking the soft limit as \eqref{eq:verticesRR-2} we have extracted the consistency relation \eqref{eq:sss-consistency} that is equivalent to what we have obtained previously as shown in Appendix~\ref{app:equiv}. As mentioned above, unlike the connected Green's function approach, we can extract the consistency relations in different momentum configurations as \eqref{eq:consistency-equil} in the equilateral one, but model dependence should be supplemented.

\subsection*{Acknowledgements}

We thank 
Daniel Chung, Toshifumi Noumi, Chang Sub Shin, Gary Shiu and Yuko Urakawa
for discussions.
MS is grateful to String Theory and Theoretical Cosmology research group, Department of Physics at the University of Wisconsin-Madison for hospitality during his visit.
JG acknowledges the support from the Korea Ministry of Education, Science and Technology, Gyeongsangbuk-Do and Pohang City for Independent Junior Research Groups at the Asia Pacific Center for Theoretical Physics. JG is also supported in part by a TJ Park Science Fellowship of POSCO TJ Park Foundation and the Basic Science Research Program through the National Research Foundation of Korea Research Grant 2016R1D1A1B03930408.
MS is supported by IBS under the project code IBS-R018-D1.

\appendix

\renewcommand{\theequation}{\Alph{section}.\arabic{equation}}

\section{Contribution of two-particle state}
\label{app:2particle}
\setcounter{equation}{0}

We first consider the successive operator product in the configuration space $\calR^2(\mathbi{x})$. We find
\begin{align}
i\big[Q,\calR^2(\mathbi{x})\big] & 
= i[Q,\calR(\mathbi{x})]\calR(\mathbi{x}) + i\calR(\mathbi{x})[Q,\calR(\mathbi{x})] 
\nonumber\\
& = \left[ -1-x^i\frac{\partial\calR(\mathbi{x})}{\partial x^i} \right] \calR(\mathbi{x}) 
+ \calR(\mathbi{x}) \left[ -1-y^i\frac{\partial\calR(\mathbi{x})}{\partial x^i} \right]
\end{align}
so that
\begin{equation}
\label{eq:2particle1}
\big\langle Q\calR^2(\mathbi{x}) \big\rangle = 
\frac{i}{2} \bigg[ x^i \left\langle \frac{\partial\calR(\mathbi{x})}{\partial x^i}\calR(\mathbi{x}) \right\rangle 
+ y^i \left\langle \calR(\mathbi{x})\frac{\partial\calR(\mathbi{x})}{\partial x^i} \right\rangle \bigg] \, .
\end{equation}
Moving to the Fourier space, the left-hand side of \eqref{eq:2particle1} becomes
\begin{equation}
\label{eq:R2-lhs}
\big\langle Q\calR(\mathbi{y})\calR^2(\mathbi{x}) \big\rangle 
= \int \frac{d^3k}{(2\pi)^3} e^{i\mathbi{k}\cdot\mathbi{x}} 
\int \frac{d^3q}{(2\pi)^3} \langle Q\calR(\mathbi{q})\calR(\mathbi{k}-\mathbi{q}) \rangle \, .
\end{equation}

For the right-hand side of \eqref{eq:2particle1} we first proceed straightforwardly to write
\begin{align}
\label{eq:R2-rhs}
& \frac{i}{2} \bigg[ x^i \left\langle \frac{\partial\calR(\mathbi{x})}{\partial x^i}\calR(\mathbi{x}) \right\rangle + x^i \left\langle \calR(\mathbi{x})\frac{\partial\calR(\mathbi{x})}{\partial x^i} \right\rangle \bigg] 
\nonumber\\
& = -3i \left\langle \int \frac{d^3k_1}{(2\pi)^3} e^{i\mathbi{k}_1\cdot\mathbi{x}} \calR(\mathbi{k}_1) 
\int \frac{d^3k_2}{(2\pi)^3} e^{i\mathbi{k}_2\cdot\mathbi{x}} \calR(\mathbi{k}_2) \right\rangle
\nonumber\\
& \quad - \frac{i}{2} \left\langle \int \frac{d^3k_1}{(2\pi)^3} e^{i\mathbi{k}_1\cdot\mathbi{x}} \Big[ \mathbi{k}_1\cdot\nabla_{\mathbi{k}_1} \calR(\mathbi{k}_1) \Big] 
\int \frac{d^3k_2}{(2\pi)^3} e^{i\mathbi{k}_2\cdot\mathbi{x}} \calR(\mathbi{k}_2) \right\rangle
\nonumber\\
& \quad - \frac{i}{2} \left\langle \int \frac{d^3k_1}{(2\pi)^3} e^{i\mathbi{k}_1\cdot\mathbi{x}} \calR(\mathbi{k}_1) 
\int \frac{d^3k_2}{(2\pi)^3} e^{i\mathbi{k}_2\cdot\mathbi{x}} \Big[ \mathbi{k}_2\cdot\nabla_{\mathbi{k}_2} \calR(\mathbi{k}_2) \Big] \right\rangle \, .
\end{align}
The first term of \eqref{eq:R2-rhs} is simply 
\begin{align}
& -3i \left\langle \int \frac{d^3k_1}{(2\pi)^3} e^{i\mathbi{k}_1\cdot\mathbi{x}} \calR(\mathbi{k}_1) 
\int \frac{d^3k_2}{(2\pi)^3} e^{i\mathbi{k}_2\cdot\mathbi{x}} \calR(\mathbi{k}_2) \right\rangle 
\nonumber\\
& = \int \frac{d^3k}{(2\pi)^3} e^{i\mathbi{k}\cdot\mathbi{x}} 
\int \frac{d^3q}{(2\pi)^3} (-3i) (2\pi)^3 \delta^{(3)}(\mathbi{k}) P_\calR(|\mathbi{k}-\mathbi{q}|) \, .
\end{align}
For the second term of \eqref{eq:R2-rhs}, we can write
\begin{align}
\label{eq:R2-rhs2nd}
& -\frac{i}{2} \left\langle \int \frac{d^3k_1}{(2\pi)^3} e^{i\mathbi{k}_1\cdot\mathbi{x}} \Big[ \mathbi{k}_1\cdot\nabla_{\mathbi{k}_1} \calR(\mathbi{k}_1) \Big] \int \frac{d^3k_2}{(2\pi)^3} e^{i\mathbi{k}_2\cdot\mathbi{x}} \calR(\mathbi{k}_2) \right\rangle
\nonumber\\
& = -\frac{i}{2} \int \frac{d^3k_1}{(2\pi)^3} \frac{d^3k_2}{(2\pi)^3} e^{i\mathbi{k}_1\cdot\mathbi{x}} e^{i\mathbi{k}_2\cdot\mathbi{x}} \mathbi{k}_1\cdot\nabla_{\mathbi{k}_1} \left\langle \calR(\mathbi{k}_1)\calR(\mathbi{k}_2) \right\rangle 
\nonumber\\
& = -\frac{i}{2} \int \frac{d^3k_1}{(2\pi)^3} \frac{d^3k_2}{(2\pi)^3} e^{i\mathbi{k}_1\cdot\mathbi{x}} e^{i\mathbi{k}_2\cdot\mathbi{x}} (2\pi)^3 \delta^{(3)}(\mathbi{k}_1+\mathbi{k}_2) (n_\calR-4) P_\calR(k_1) \, ,
\end{align}
where we have used the scaling relation of the power spectrum $k^3P_\calR(k) \propto k^{n_\calR-1}$. Furthermore, since the power spectrum is dependent only on $|\mathbi{k}_1| = |\mathbi{k}_2|$ that is imposed by the delta function, we can replace the argument of the power spectrum with $k_2$: $P_\calR(k_1) = P_\calR(k_2)$. Then we find
\begin{equation}
\text{\eqref{eq:R2-rhs2nd}} = \int \frac{d^3k}{(2\pi)^3} e^{i\mathbi{k}\cdot\mathbi{x}} \int \frac{d^3q}{(2\pi)^3} \frac{-i}{2} (2\pi)^3 \delta^{(3)}(\mathbi{k}) (n_\calR-4) P_\calR(|\mathbi{k}-\mathbi{q}|) \, .
\end{equation}
We can follow similar steps for the third term of \eqref{eq:R2-rhs}. Then finally \eqref{eq:R2-rhs} becomes
\begin{equation}
\text{\eqref{eq:R2-rhs}} = \int \frac{d^3k}{(2\pi)^3} e^{i\mathbi{k}\cdot\mathbi{x}} 
\int \frac{d^3q}{(2\pi)^3} i(1-n_\calR) (2\pi)^3 \delta^{(3)}(\mathbi{k}) P_\calR(|\mathbi{k}-\mathbi{q}|) \, .
\end{equation}
This is to be equated with \eqref{eq:R2-lhs}.

Now we consider the contribution of the two-particle state $\left|\calR^2,\mathbi{q}\right\rangle$, i.e. the first line of \eqref{eq:R2state-result}. Let us write, as we did for the one-particle state, 
\begin{equation}
\left|\calR^2,\mathbi{k}\right\rangle 
= c_2(\mathbi{k})\left( \calR^2 \right)(\mathbi{k})|\Omega\rangle
= c_2(\mathbi{k}) \int \frac{d^3q}{(2\pi)^3} 
\calR(\mathbi{q})\calR(\mathbi{k}-\mathbi{q}) |\Omega\rangle \, .
\end{equation}
From $\left\langle \left( \calR^2 \right)(\mathbi{q}) \right| = \langle\Omega| c_2^*(\mathbi{q}) \left( \calR^2 \right)^\dag(\mathbi{q}) = \langle\Omega| c_2^*(\mathbi{q}) \left( \calR^2 \right)(-\mathbi{q})$, demanding the normalization condition gives
\begin{align}
\left\langle \calR^2,\mathbi{q} \big| \calR^2,\mathbi{k}\right\rangle 
& = (2\pi)^3 \delta^{(3)}(\mathbi{k}-\mathbi{q})
\nonumber\\
& = c_2^*(\mathbi{q})c_2(\mathbi{k}) 
\left\langle \Omega \left| \left( \calR^2 \right)^\dag(\mathbi{q})
\left( \calR^2 \right)(\mathbi{k}) \right| \Omega \right\rangle
\nonumber\\
& = (2\pi)^3 \delta^{(3)}(\mathbi{k}-\mathbi{q})
c_2^*(\mathbi{q})c_2(\mathbi{k}) \int \frac{d^3p_1}{(2\pi)^3} \frac{d^3p_2}{(2\pi)^3}
T_\calR(-\mathbi{p}_1,-\mathbi{q}+\mathbi{p}_1,\mathbi{p}_2,\mathbi{k}-\mathbi{p}_2) \, ,
\end{align}
with the trispectrum being defined by $\langle\calR(\mathbi{k}_1)\cdots\calR(\mathbi{k}_4)\rangle \equiv (2\pi)^3\delta^{(3)}\left(\sum_{i=1}^4\mathbi{k}_i\right)T_\calR(\mathbi{k}_1,\mathbi{k}_2,\mathbi{k}_3,\mathbi{k}_4)$. Since the delta function imposes $\mathbi{k}=\mathbi{q}$, we obtain
\begin{equation}
c_2(\mathbi{q}) = \left[ \int \frac{d^3p_1}{(2\pi)^3} \frac{d^3p_2}{(2\pi)^3} T_\calR(-\mathbi{p}_1,-\mathbi{q}+\mathbi{p}_1,\mathbi{p}_2,\mathbi{q}-\mathbi{p}_2) \right]^{-1/2} \, ,
\end{equation}
Then we find
\begin{align}
\left\langle\Omega\left| Q \right|\calR^2,\mathbi{q}\right\rangle & = \int \frac{d^3p}{(2\pi)^3} c_2(\mathbi{q}) 
\langle \Omega | Q \calR(\mathbi{p})\calR(\mathbi{q}-\mathbi{p}) |\Omega\rangle
\nonumber\\
& = \int \frac{d^3p}{(2\pi)^3} c_2(\mathbi{q}) i(1-n_\calR) (2\pi)^3 \delta^{(3)}(\mathbi{q}) P_\calR(|\mathbi{q}-\mathbi{p}|) \, ,
\\
\left\langle \Omega \left| \calR(\mathbi{k}_1) \cdots \calR(\mathbi{k}_n) \right| \calR^2,\mathbi{q} \right\rangle & = \int \frac{d^3p}{(2\pi)^3} c_2(\mathbi{q}) \langle \calR(\mathbi{k}_1) \cdots \calR(\mathbi{k}_n) \calR(\mathbi{p})\calR(\mathbi{q}-\mathbi{p}) \rangle \, .
\end{align}
Using the symmetry of the correlation functions under the exchange of momenta and parity, we finally arrive at \eqref{eq:R2state-result}.

\section{Equivalence to connected Green's function approach}
\label{app:equiv}
\setcounter{equation}{0}

In order to check the equivalence between the 1PI Slavnov-Taylor identities and \eqref{eq:multiRRR2-3}, we make use of the standard expressions for connected Green's functions written in terms of the 1PI Green's functions. For two-point connected Green's functions, they are given by
\begin{align}
\label{Eq:1PIrel1}
-P_\calR & = ( \Gamma_{\calR\calR}-\Gamma_{\calR a}\Gamma_{ab}^{-1} \Gamma_{b\calR})^{-1} \, ,
\\
\label{Eq:1PIrel2}
-\widehat{P}_{\calR a} & = \Gamma_{\calR\calR}^{-1} \Gamma_{\calR b} P_{ab}=P_\calR\Gamma_{\calR b}\Gamma_{ba}^{-1} \, ,
\\
\label{Eq:1PIrel3}
-\widehat{P}_{a\calR} & = P_{ab} \Gamma_{b\calR} \Gamma_{\calR\calR}^{-1}= \Gamma_{ab}^{-1}\Gamma_{b\calR}P_\calR \, ,
\\
\label{Eq:1PIrel4}
-\widehat{P}_{ab} & = (\Gamma_{ab}-\Gamma_{a\calR}\Gamma_{\calR\calR}^{-1}\Gamma_{Rb})^{-1} \, .
\end{align}
Here, the connected Green's functions with hats above mean they are given by $\widehat{\varphi}_\perp^a$ rather than $\varphi_\perp^a$. The three-point connected Green's functions are given by
\begin{align}
\label{Eq:BRRR}
B_\calR(\mathbi{q},\mathbi{k},-\mathbi{k}-\mathbi{q}) & =
P_\calR(q)P_\calR(k)P_\calR(|\mathbi{k}+\mathbi{q}|) 
\Gamma_{\calR\calR\calR}(\mathbi{q},\mathbi{k},-\mathbi{k}-\mathbi{q})
\nonumber\\
& \quad + P_\calR(q)P_\calR(k)\widehat{P}_{a\calR}(|\mathbi{k}+\mathbi{q}|)
\Gamma_{\calR\calR a}(\mathbi{q},\mathbi{k},-\mathbi{k}-\mathbi{q}) 
+ \text{2 perm}
\nonumber\\
& \quad + \widehat{P}_{a\calR}(q)\widehat{P}_{b\calR}(k)P_\calR(|\mathbi{k}+\mathbi{q}|)
\Gamma_{ab\calR}(\mathbi{q},\mathbi{k},-\mathbi{k}-\mathbi{q}) + \text{2 perm}
\nonumber\\
& \quad + \widehat{P}_{a\calR}(q) \widehat{P}_{b\calR}(k) \widehat{P}_{c\calR}(|\mathbi{k}+\mathbi{q}|)
\Gamma_{abc}(\mathbi{q},\mathbi{k},-\mathbi{k}-\mathbi{q}) \, ,
\\
\label{Eq:BaRR}
\widehat{B}_{a\calR\calR}(\mathbi{q},\mathbi{k},-\mathbi{k}-\mathbi{q}) & =
\widehat{P}_{a\calR}(q)P_\calR(k)P_\calR(|\mathbi{k}+\mathbi{q}|)
\Gamma_{\calR\calR\calR}(\mathbi{q},\mathbi{k},-\mathbi{k}-\mathbi{q}) 
\nonumber\\
& \quad + \widehat{P}_{a\calR}(q)P_\calR(k)\widehat{P}_{\calR b}(|\mathbi{k}+\mathbi{q}|)
\Gamma_{\calR\calR b}(\mathbi{q},\mathbi{k},-\mathbi{k}-\mathbi{q})
+ (k \leftrightarrow \mathbi{k}+\mathbi{q})
\nonumber\\
& \quad + \widehat{P}_{ab}(q)P_\calR(k)P_\calR(|\mathbi{k}+\mathbi{q}|)
\Gamma_{b\calR\calR}(\mathbi{q},\mathbi{k},-\mathbi{k}-\mathbi{q}) 
\nonumber\\
& \quad + \widehat{P}_{ab}(q)P_\calR(k)\widehat{P}_{\calR c}(|\mathbi{k}+\mathbi{q}|)
\Gamma_{b\calR c}(\mathbi{q},\mathbi{k},-\mathbi{k}-\mathbi{q}) 
+ (k \leftrightarrow \mathbi{k}+\mathbi{q})
\nonumber\\
& \quad + \widehat{P}_{a\calR}(q)\widehat{P}_{b\calR}(k)\widehat{P}_{c\calR}(|\mathbi{k}+\mathbi{q}|)
\Gamma_{\calR bc}(\mathbi{q},\mathbi{k},-\mathbi{k}-\mathbi{q}) 
\nonumber\\
& \quad + \widehat{P}_{ab}(q)\widehat{P}_{c\calR}(k)\widehat{P}_{d\calR}(|\mathbi{k}+\mathbi{q}|)
\Gamma_{bcd}(\mathbi{q},\mathbi{k},-\mathbi{k}-\mathbi{q}) \, .
\end{align}

With $P_a \gg P_{ab}$ for $a \ne b$, $\widehat{\varphi}_\perp^a$ are related to $\varphi_\perp^a$ through
\begin{equation}
\label{eq:orthogonalop}
\widehat{\varphi}_\perp^a \approx \varphi_\perp^a - \sum_{b \ne a} \frac{P_{ab}}{P_b}\widehat{\varphi}_\perp^b
\end{equation}
as implied in \eqref{eq:orthogonalstate}. In this case, $\widehat{P}_{ab}$ is diagonalized to be proportional to $\delta_{ab}$ perturbatively, satisfying
\begin{equation}
\begin{split}
\widehat{P}_a & \approx P_a - \sum_{b \ne a}\frac{P_{ab}^2}{P_b} \, ,
\\
\widehat{B}_{\calR\calR a} & \approx B_{\calR\calR a} - \sum_{a \ne b} \frac{P_{ab}}{P_b} B_{\calR\calR b} \, ,
\end{split}
\end{equation}
and so on. Substituting these into \eqref{eq:multiRRR2-3} and using \eqref{Eq:1PIrel1} and \eqref{Eq:1PIrel2}, we find that \eqref{eq:multiRRR2-3} becomes simply
\begin{equation}
(1-n_\calR)P_\calR(k) = -\Gamma_{\calR\calR}(q) B_\calR(k_1, k_2, q) - \Gamma_{\calR a}(q) \widehat{B}_{\calR\calR a}(k_1, k_2, q) \, ,\label{Eq:diagconsist}
\end{equation}
where we have omitted $\lim_{q\to0}$ for notational simplicity. Then using \eqref{Eq:BRRR} and \eqref{Eq:BaRR}, the right-hand side of \eqref{Eq:diagconsist} becomes
\begin{align}
\label{eq:1pi-cgf-rhs}
& -\Gamma_{\calR\calR}(q) B_\calR(k_1, k_2, q) - \Gamma_{\calR a}(q) \widehat{B}_{\calR\calR a}(k_1, k_2, q) 
\nonumber\\
& = P_\calR(k)P_\calR(|\mathbi{k}+\mathbi{q}|)\Gamma_{\calR\calR\calR}(\mathbi{q},\mathbi{k},-\mathbi{k}-\mathbi{q})
+ P_\calR(k)\widehat{P}_{b\calR}(|\mathbi{k}+\mathbi{q}|)\Gamma_{\calR\calR b}(\mathbi{q},\mathbi{k},-\mathbi{k}-\mathbi{q})
\nonumber\\
& \quad +\widehat{P}_{b\calR}(k)P_{\calR}(|\mathbi{k}+\mathbi{q}|)\Gamma_{\calR b \calR}(\mathbi{q},\mathbi{k},-\mathbi{k}-\mathbi{q})
+ \widehat{P}_{b\calR}(k)\widehat{P}_{c\calR}(|\mathbi{k}+\mathbi{q}|)\Gamma_{\calR bc}(\mathbi{q},\mathbi{k},-\mathbi{k}-\mathbi{q}) \, .
\end{align}
Further, now making use of the trace parts of the Slavnov-Taylor identities derived in Section~\ref{subsec:1PIconsist}, we further find
\begin{align}
\label{eq:1pi-cgf-rhs2}
\eqref{eq:1pi-cgf-rhs} & = 3 \Big[ 
P_\calR(k)P_\calR(k)\Gamma_{\calR\calR}(k)
+P_\calR(k)\widehat{P}_{b\calR}(k)\Gamma_{\calR b}(k)
+\widehat{P}_{b\calR}(k)P_{\calR}(k)\Gamma_{b \calR}(k)
+\widehat{P}_{b\calR}(k)\widehat{P}_{c\calR}(k)\Gamma_{bc}(k) \Big]
\nonumber\\
& \quad - \mathbi{k} \cdot \Big[
P_\calR(k)P_\calR(k)\nabla_{\mathbi{k}}\Gamma_{\calR\calR}(k)
+ 2P_\calR(k)\widehat{P}_{b\calR}(k)\nabla_{\mathbi{k}}\Gamma_{\calR b}(k)
+\widehat{P}_{b\calR}(k)\widehat{P}_{c\calR}(k)\nabla_{\mathbi{k}}\Gamma_{bc}(k)
\Big] \, .
\end{align}
By using \eqref{Eq:1PIrel1}, \eqref{Eq:1PIrel2}, \eqref{Eq:1PIrel3} and \eqref{Eq:1PIrel4}, we can trivially find 
\begin{align}
\eqref{eq:1pi-cgf-rhs2} & = -3P_\calR(k) + \mathbi{k} \cdot \nabla_{\mathbi{k}}
\Big[ \Gamma_{\calR\calR}^{-1} - \Gamma_{\calR\calR}^{-1} \Gamma_{\calR a} \widehat{P}_{ab} \Gamma_{b\calR} \Gamma_{\calR\calR}^{-1}
\Big] 
\nonumber\\
& = -(3+\mathbi{k}\cdot\nabla_{\mathbi{k}}) P_\calR(k) \, .
\end{align}
Thus the 1PI Green's function Slavnov-Taylor identities exactly reproduces the consistency relation, proving the equivalence between the two approaches.

\end{document}